# Strange electrical transport: Colossal magnetoresistance via avoiding fully polarized magnetization in ferrimagnetic insulator $Mn_3Si_2Te_6$

Yifei Ni[1], Hengdi Zhao[1], Yu Zhang[1], Bing Hu[1,2] Itamar Kimchi[3] and Gang Cao[1]*

[1]Department of Physics, University of Colorado at Boulder, Boulder, CO 80309, USA

[2]School of Mathematics and Physics, North China Electric Power University, Beijing 102206, China

[3]School of Physics, Georgia Institute of Technology, Atlanta, GA 30332, USA

Colossal magnetoresistance is of great fundamental and technological significance and exists mostly in the manganites and a few other materials. Here we report colossal magnetoresistance that is starkly different from that in all other materials. The stoichiometric $Mn_3Si_2Te_6$ is an insulator featuring a ferrimagnetic transition at 78 K. The resistivity drops by 7 orders of magnitude with an applied magnetic field above 9 Tesla, leading to an insulator-metal transition at up to 130 K. However, the colossal magnetoresistance occurs only when the magnetic field is applied along the magnetic hard axis and is surprisingly absent when the magnetic field is applied along the magnetic easy axis where magnetization is fully saturated. The anisotropy field separating the easy and hard axes is 13 Tesla, unexpected for the Mn ions with nominally negligible orbital momentum and spin-orbit interactions. Double exchange and Jahn-Teller distortions that drive the hole-doped manganites do not exist in $Mn_3Si_2Te_6$. The phenomena fit no existing models, suggesting a unique, intriguing type of electrical transport.

*gang.cao@colorado.edu

Colossal magnetoresistance (CMR) was first observed in the hole-doped perovskite manganites in the early 1990's. This discovery was followed by an explosion of interest in these materials. In the following decades, the extensive efforts have resulted in comprehensive insights into the rich physics of this class of materials [1-16]. The hole-doped manganites, such as $La_{1-x}Ca_xMnO_3$, feature mixed oxidation states of $Mn^{3+}$ and $Mn^{4+}$, a key element of this class of CMR materials. In essence, the concurrent magnetic and insulator-metal transitions in these materials arise from a combined effect of double exchange, which dictates magnetism, and Jahn-Teller distortions, which drives electrical transport [13-16]. An important exception was later found in the pyrochlore $Tl_2Mn_2O_7$ having the nominal oxidation state of $Mn^{4+}$, without double exchange and Jahn-Teller polarons [17]. This material undergoes a simultaneous ferromagnetic and insulator-metal transition at 135 K, resulting in CMR near the transition temperature [17]. A theoretical model developed for the pyrochlore manganite attributes the phenomena to magnetic polarons forming above the Curie temperature on condition that the carrier density is sufficiently low [18]. This model is also used to explain CMR recently discovered in the antiferromagnetic $Eu_5In_2Sb_6$ [19]. Negative magnetoresistivity (MR) is also predicted in topological Weyl/Dirac semimetals, arising from the chiral anomaly $\vec{E} \cdot \vec{B}$ term, thus, relying on an electrical current parallel to magnetic fields [20]. It has become increasingly clear that CMR, a phenomenon of great fundamental and technological significance, is far from fully explored and understood.

Here we report a type of CMR in the stoichiometric chalcogenide $Mn_3Si_2Te_6$ that happens via avoiding a fully polarized magnetization, distinguishing from that observed previously in other materials. The three-dimensional $Mn_3Si_2Te_6$ with the $Mn^{2+}(3d^5)$ ions is a ferrimagnetic insulator with a ferrimagnetic Néel temperature $T_C$ = 78 K [21-25]. This work reveals that the electrical resistivity drastically reduces by *seven orders of magnitude or 99.99999% (1 - $10^{-7}$ = 1 - 0.0000001*

= 0.9999999 = 99.99999%) at an applied magnetic field, H, above 9 T, inducing an insulator-metal transition, $T_{IM}$, at up to 130 K (Note that MR is defined as $[\rho(H)-\rho(0)]/\rho(0)$). The reduction of resistivity is arguably among the largest values of CMR reported thus far. However, starkly different from CMR in other materials, *the observed CMR occurs only when H is applied along the magnetic hard axis, the c axis, and is absent when H is applied along the magnetic easy axis which lies within the ab plane.* The *ab*-plane magnetization readily and fully saturates at $\mu_o H \geq$ 0.05 T with a saturation magnetization, $M_s$, of 1.56 $\mu_B$/Mn whereas the *c*-axis magnetization approaches this value only when $\mu_o H \geq$ 13 T. Such an unusually strong magnetic anisotropy of 13 T is unexpected for the 3d ions with nominally quenched orbital momentum and negligible spin-orbit interactions (SOI). Moreover, our Hall effect data indicate a low carrier density ranging from $10^{23}$/m$^3$ to $10^{24}$/m$^3$. This type of CMR is both new and intriguing and calls for urgent attention. (For experimental details see Ref. [26].)

$Mn_3Si_2Te_6$ has been known since 1981 [21, 22]. It has drawn more attention in recent years because of its apparent relevance to van der Waals materials [23-25]. In this material, inherent frustration due to competing exchange interactions prevents a long-range order from setting in until the temperature is lowered to $T_C$ = 78 K [23]. Results of diffuse magnetic scattering suggest short-range spin correlations existing well above $T_C$, possibly persisting up to 330 K [23]. Spin fluctuations thus may have important implications in the phenomena discussed below.

*Crystal Structure* $Mn_3Si_2Te_6$ adopts a trigonal structure having space group no. 163 [23]. Our single-crystal X-ray diffraction data as a function of temperature from 80 K to 300 K are consistent with the reported results. In essence, $Mn_3Si_2Te_6$ consists of $MnTe_6$ octahedra edge-sharing within the *ab* plane and face-sharing along the *c* axis. Remarkably, the bond distance Mn1-Mn1 of two neighboring *edge-sharing* $MnTe_6$ octahedra in the *ab* plane is 4.0520 Å whereas the bond distance

Mn1-Mn2 of two neighboring *face-sharing* MnTe$_6$ octahedra along the *c* axis is much shorter, merely 3.5200 Å at 80 K, rendering more consequential exchange interactions between Mn1 and Mn2 (**Fig.1a**). The magnetic spins are ferromagnetically aligned within the *ab* plane and antiferromagnetically aligned along the *c* axis below T$_C$ (**Fig.1b**). No magnetic canting is discerned [23]. The temperature dependence of the lattice parameters indicates that the *c* axis shrinks at a faster rate than the *a* axis from 300 K to 80 K (**Fig.1c**). No crystal structural change is observed. Mn$_3$Si$_2$Te$_6$ is clearly a robust three-dimensional lattice. The crystals are of 1-2 millimeters in size (**Fig.1d**).

*Magnetic Properties* The magnetization, M, is measured as functions of magnetic field H and temperature, T. The ferrimagnetic transition T$_C$ = 78 K broadens and shifts to higher temperatures as H increases (**Figs.2a-2b**). Moreover, a well-defined magnetic anomaly also occurs at T* = 330 K (**Inset** in **Fig.2a**), suggesting a short-range order due to inherent frustration [23]. Indeed, the temperature dependence of the *ab*-plane magnetization, M$_{ab}$, between T$_C$ and T* noticeably deviates from the Curie-Weiss law, indicating an absence of an anticipated paramagnetic state (**Inset** in **Fig.2a**).

The magnetic easy axis lies within the *ab* plane, where M$_{ab}$ fully saturates at $\mu_oH > 0.05$ T and T = 10 K, resulting in M$_s$ = 1.56 $\mu_B$/Mn (**Fig.2c**). In contrast, the *c*-axis magnetization M$_c$ approaches this value only when $\mu_oH \geq 13$ T. Note that M$_c$ remains smaller than M$_{ab}$. Such a large anisotropy field of 13 T or ~ 10$^8$ A/m is unexpected for the 3d ions where both orbital momentum and SOI are nominally negligible. In general, magnetic anisotropy occurs because the crystal field stabilizes a preferred orbital that, via SOI, aligns the spin along a preferred crystallographic direction. However, the observed anisotropy field is surprisingly consistent with an ab-initio calculation [23], predicting an anisotropy field of 13 T. The calculation suggests that the orbital moment is 0.037 $\mu_B$/Mn1 and −0.048 $\mu_B$/Mn2, respectively [23]. Apparently, the orbital moment and SOI, however small, are surprisingly consequential in this material. Note that the observed $M_s$ (= 1.56 $\mu_B$/Mn) is small for the Mn$^{2+}$(3d$^5$) ion in which the Hund's rules dictate a high spin state S = 5/2 or possibly an intermediate spin state S=3/2.

*Transport Properties* Mn$_3$Si$_2$Te$_6$ has an insulating ground state at ambient conditions (**Fig.3a**). The *ab*-plane resistivity, $\rho_{ab}$, rises by 10$^7$ as T decreases from 380 K to 3 K. A brief drop in $\rho_{ab}$ below $T_C$ is due to the reduction of spin scattering as the ferrimagnetic state sets in. $\rho_{ab}$ rises again and rapidly below 60 K, reaching 3 x 10$^6$ Ω cm at 3 K, and becomes unmeasurably too high below 3K. The temperature dependence of $\rho_{ab}$ below $T_C$ strongly deviates from an activation law, ruling out thermal activation as an origin of the insulating state. In addition, $\rho_{ab}$ exhibits a pronounced slope change near the magnetic anomaly T* = 330 K (**Inset** in **Fig.3a**), which diminishes upon application of magnetic field, hinting a suppressing of spin fluctuations.

Clearly, the resistivity is extremely sensitive to application of magnetic field. As demonstrated in **Fig.3a**, $\rho_{ab}$ drastically reduces by *seven orders of magnitude* at low temperatures, leading to an insulator-metal transition $T_{IM}$ when H||c, e.g., $T_{IM} \approx$ 130 K for $\mu_oH_{||c}$=14 T.

The *c*-axis resistivity $\rho_c$ responds to H in a similar but less dramatic manner (**Fig.3b**). However, unlike $\rho_{ab}$, $\rho_c$ forms a pronounced valley between 30 K and 60 K that reaches 10$^{-2}$ Ω cm before climbing back to a higher value. Note that $\rho_c < \rho_{ab}$ (**Inset** in **Fig.3b**), implying the importance of the shorter Mn1-Mn2 bond distance.

*The resistivity is extremely sensitive to the direction of H in an unanticipated manner*. As shown in **Figs. 4a,** when H||ab, $\rho_{ab}$ drops by mere 20% at 14 T (red solid lines). This is strikingly unusual because the magnetic easy axis lies within the *ab* plane, where $M_{ab}$ readily and fully saturates at 0.05 T, reaching $M_s$ (red dashed line). In sharp contrast, when H||c, the magnetic hard

axis, $\rho_{ab}$ drops by 99.99999% at $\mu_oH_{\|c} \geq 9$ T at which $M_c < M_s$ (**Fig.4b**). Note that $\rho_{ab}$ already decreases by 97.50% at $\mu_oH_{\|c} = 3$ T at which $M_c = 0.77$ $\mu_B$/Mn $< 0.5M_s$ (blue dashed line in **Fig.4a**). Even at $T > T_C$, the absolute value of the negative MR remains large, e.g., 85% at 120 K (**Fig.4c**). In short, *the CMR is not coupled with the fully polarized magnetization $M_s$ and emerges only when $M_s$ is avoided, contradicting CMR in other materials where magnetic polarization is essential.*

The angular dependence of $\rho_{ab}$ at 14 T at various temperatures provides more insight into this behavior. A polar plot generated based on the data is shown in **Fig.4d**. The angle, $\theta$, measures the angle between H and the *c* axis, e.g., $\theta = 0°$ for H$\|$c and $\theta = 90°$ for H$\|$ab (**Inset** in **Fig.4d**). $\rho_{ab}$ as a function of $\theta$ forms elongated lobes pointing to $\theta=90°$ or $270°$ at 10 K and 30 K. It virtualizes the extraordinary anisotropy of $\rho_{ab}$ that is smallest when H is parallel to the magnetic hard axis ($\theta = 0°$ or $180°$) and largest when H is parallel to the magnetic easy axis ($\theta = 90°$ or $270°$). The anisotropy weakens but is still visible at 120 K, well above $T_C$ (= 78 K), implying once again that the state above $T_C$ involves substantial short-range correlations and is not a simple paramagnet.

This is also supported by the results of Hall effect (**Fig.5a**). The Hall resistivity, $\rho_{xy}$, as a function of H$\|$c exhibits a pronounced peak in vicinity of 0.35 T for two representative temperatures below $T_C$, indicating the anticipated anomalous Hall effect (AHE). However, this peak persists, though broadened, above $T_C$, e.g., 120 K. This is unanticipated for a paramagnet where $\rho_{xy}$ changes linearly with H. The linearity is eventually recovered at 200 K (**Lower Inset**)

The carrier density, n, is retrieved from the linear portion of $\rho_{xy}$(H) in a range of 11 – 14 T, well beyond the AHE peak, with subtraction of the residual AHE contribution. The assumption of a one-band structure might be overly simplistic, but it offers a useful estimate of n in this case. As shown in **Fig.5b**, n, ranging from $10^{23}$/m$^3$ to $10^{24}$/m$^3$, closely tracks $\rho_{ab}$(T) at 14 T (red dashed line). The sharp rise of n below 130 K indicates rapid delocalization of holes, leading to the metallic state.

The low n is consistent with the results from a low-field scaling relation of $[\rho(0)-\rho(H)]/\rho(0) = C\,(M/M_s)^2$ where $C$ is a scaling constant proportional to $1/n^{2/3}$ (**Inset** in **Fig.4c**). The value of $C$ is reportedly smaller than 10 for hole-doped manganites [7], 15 for $Tl_2Mn_2O_7$ [17] and 50 for $Eu_5In_2Sb_6$ [19]. For $Mn_3Si_2Te_6$, $C=105$, implying that the carrier density is indeed particularly low.

The heat capacity, C(T), plotted as C(T)/T vs $T^2$ (**Fig.5c**) shows a large linear contribution of C(T) ~ T with C(T)/T = 23 mJ/mole $K^2$ at T = 0 and H = 0. This is anticipated for a correlated metal with a high density of electronic states but unexpected for an insulator such as $Mn_3Si_2Te_6$. This linear term is suppressed by H (consistent with earlier reports [23]) and reduces to zero at $\mu_oH_{\parallel c}$ = 14 T when $Mn_3Si_2Te_6$ becomes a metal featuring the drastically enhanced n and conductivity below $T_{IM}$ (**Fig.5b**). Clearly, the states that produce the linear term must lie within the magnetic sector and cannot be the conduction electrons.

The large linear term suggests gapless or critical magnetic fluctuations such as those arising from a coexisting or proximate critical phase, e.g., a spinon Luttinger liquid or a spinon-Fermi-surface quantum spin liquid [27-29]. Indeed, effective Luttinger liquids could emerge from the strong c-axis $J_1$ bonds, whereas spin liquids could occur due to sufficient frustration, which is evidenced by the suppressed $T_C$, the absence of Curie-Weiss behavior up to T ≈ 4$T_C$, and the competing exchange interactions with $J_{1,2,3}$ showing $|J_3| > |J_2|$ [23]. Localized electrons could contribute a linear term susceptible to field suppression only via strong spin correlations leading again to gapless spinful excitations. Since the ab-plane spins are nearly fully polarized it is natural to assume that the critical magnetic fluctuations are associated with the c-axis spin degree of freedom. As such, the reduced density of states of the critical fluctuations (**Inset** in **Fig.5c**) corresponds to the slow spin polarization when H∥c (**Fig.2c**). Note that these critical fluctuations are unrelated to critical-scaling curves near $T_C$ [24] or to ferrimagnetic Goldstone modes.

The phenomena reported here cannot be explained by any existing models. However, the model developed for the ferromagnetic pyrochlore manganites [18] may provide a starting point for an eventual understanding. This model suggests that the ultra-low-density carriers can form magnetic polarons dressed by mean-field ferromagnetic spin fluctuations in an intermediate temperature regime above $T_C$, and CMR emerges via suppressing the spin fluctuations [18]. In $Mn_3Si_2Te_6$, the presence of magnetic polarons is evidenced by the absence of the Curie-Weiss behavior and the persisting CMR and AHE above $T_C$. The magnetic polarons are now mostly dressed by the critical magnetic fluctuations associated with the *c*-axis spin degree of freedom. Suppressing the critical fluctuations can produce negative MR across the wide range of temperatures where the fluctuations are present. However, the key observation – that the CMR occurs only when fully polarized magnetization is avoided – is not an outcome of the model and is not captured by any other existing models. Clearly, the CMR reported here is fundamentally different from that in other materials, providing a new direction for studying colossal magnetoresistance and its applications.

**Acknowledgement** This work is supported by NSF via grant DMR 1903888. GC is thankful to Peter Riseborough, Minhyea Lee, Dmitry Reznik, Dan Dessau and Feng Ye for useful discussions.

**Figure captions**

**Fig.1. Crystal structure (a)** The crystal structure highlighting the three-dimensional nature and bond distances for Mn1-Mn1 and Mn1-Mn2 marked by the white dashed lines. The third-nearest-neighbor interaction denoted by the diagonal dashed line proves consequential [23]. **(b)** The magnetic structure based on Ref. [23]. **(c)** The temperature dependence of the *a* and *c* axis, and the relative changes in the *a* and *c* axis against the values at 300 K (right scale). **(d)** A crystal sample showing the *ab* plane.

**Fig.2. Magnetic properties** The temperature dependence of **(a)** the *ab*-plane magnetization $M_{ab}$ and **(b)** the *c*-axis magnetization $M_c$ at various magnetic fields. **Inset** in (a): $M_{ab}$ highlighting the correlated behavior at $T_C < T < T^*$. **Inset** in (b): $M_{ab}(T)$ and $M_c(T)$ at 14 T. **(c)** The isothermal magnetization $M_{ab}$ and $M_c$ at 10 K up to 14 T highlighting the anisotropy field. Note that $M_{ab}$ saturates at $\mu_oH = 0.05$ T and $M_c$ approaches $M_s$ at $\mu_oH \geq 13$ T.

**Fig.3. Transport properties** The temperature dependence of **(a)** the *ab*-plane resistivity $\rho_{ab}$ and **(b)** the *c*-axis resistivity $\rho_c$ at various magnetic fields applied along the *c* axis. Note that $\rho_{ab}$ drops by seven orders of magnitude or 99.99999% at low temperatures and the $T_{IM}$ shifts up to 130 K at 14 T. **Inset** in (a) $\rho_{ab}$ highlighting anomaly corresponding to T*. **Inset** in (b) showing $\rho_{ab} > \rho_c$.

**Fig.4. Anisotropy and CMR below and above $T_C$** The magnetic field dependence at 10 K of **(a)** $\rho_{ab}$ (H||ab) and $\rho_{ab}$ (H||c) and $M_{ab}$ and $M_c$ (dashed lines, right scale), and **(b)** the corresponding $\Delta\rho_{ab}/\rho_{ab}$(H||ab) and $\Delta\rho_{ab}/\rho_{ab}$(H||c), where $\Delta\rho_{ab}/\rho_{ab}=[\rho_{ab}(H)-\rho_{ab}(0)]/\rho_{ab}(0)$. Note that $\Delta\rho_{ab}/\rho_{ab}$(H||ab) is ~20% at $\mu_oH_{||ab} = 14$ T and $M_{ab} = M_s$ whereas $\Delta\rho_{ab}/\rho_{ab}$(H||c) is already 97.500% at $\mu_oH_{||c} = 3$ T at which $M_c = 0.77$ $\mu_B$/Mn $< 0.5M_s$ and 99.99999% at $\mu_oH_{||c} > 9$ T at which $M_c < M_s$. **(c)** The field dependence of $\Delta\rho_{ab}/\rho_{ab}$(H||c) at 5 K, 30 K and 120 K. Note that MR is still 85% at 120 K. **Inset:** The low-field scaling plot of $\Delta\rho_{ab}/\rho_{ab}$(H||c) $= C\,(M_c/M_s)^2$ for $\mu_oH_{||c} < 3.2$ T and 80 K > $T_C$, yielding

$C = 105$ (dashed line). **(d)** The angular dependence of $\rho_{ab}$ at 14 T and various temperatures. **Inset:** The angle $\theta$ measures the angle between H and the $c$ axis.

**Fig.5. Hall effect, carrier density and heat capacity (a)** The field dependence of the Hall resistivity $\rho_{xy}$ at representative temperatures. Note the AHE persisting up to $T > T_C$. **Upper inset:** The Hall sample with electrical leads and H pointing out of the page. **Lower inset:** $\rho_{xy}(H)$ at 200 K, showing the linear H dependence. **(b)** The temperature dependence of the carrier density n and $\rho_{ab}$ at $\mu_o H_{\|c} = 14$ T (red dashed line, right scale). Note that n closely tracks $\rho_{ab}$. **(c)** The heat capacity $C(T)$ at zero field (blue) and $\mu_o H_{\|c} = 14$ T (red) plotted as $C(T)/T$ vs $T^2$. Note that the linear-T term due to the critical spin fluctuations is suppressed by H (see **Inset**).

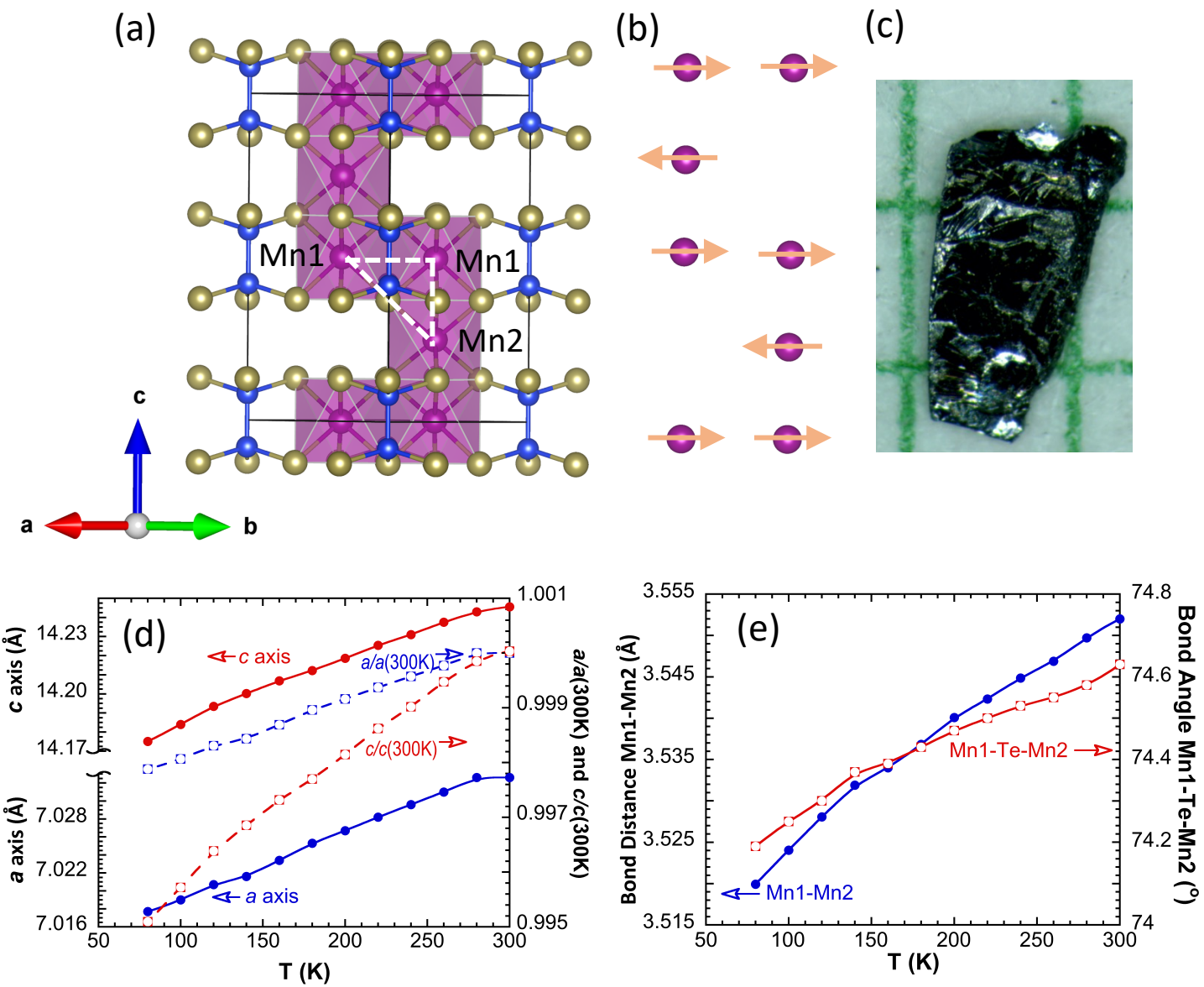

Figure 1

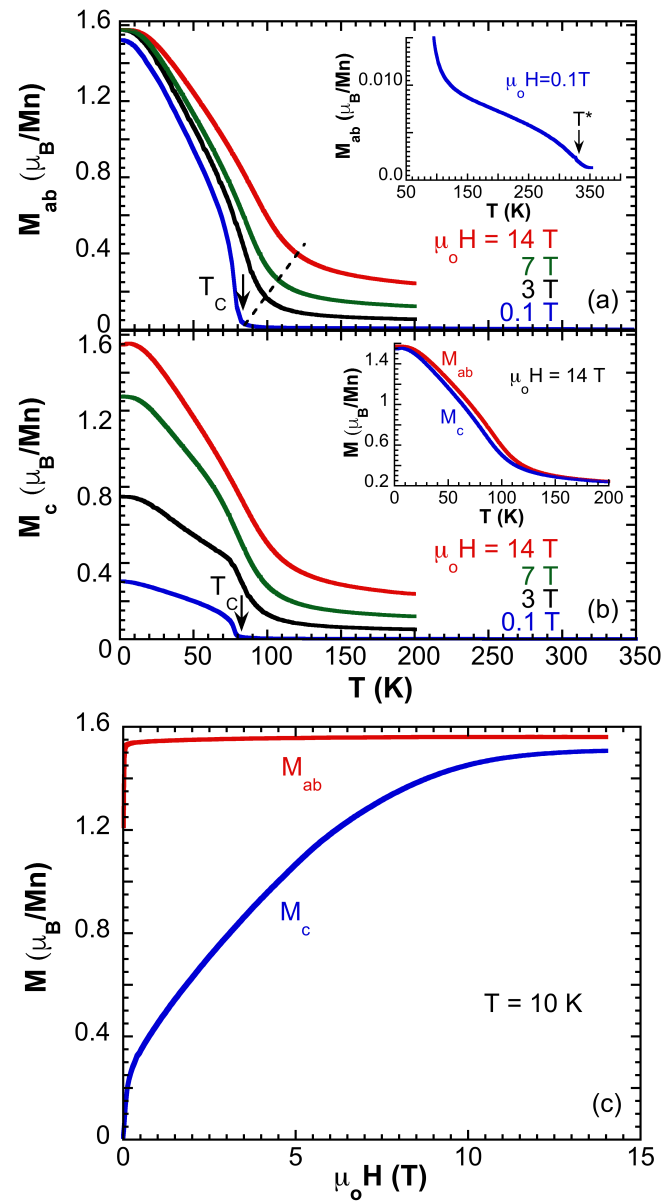

Figure 2

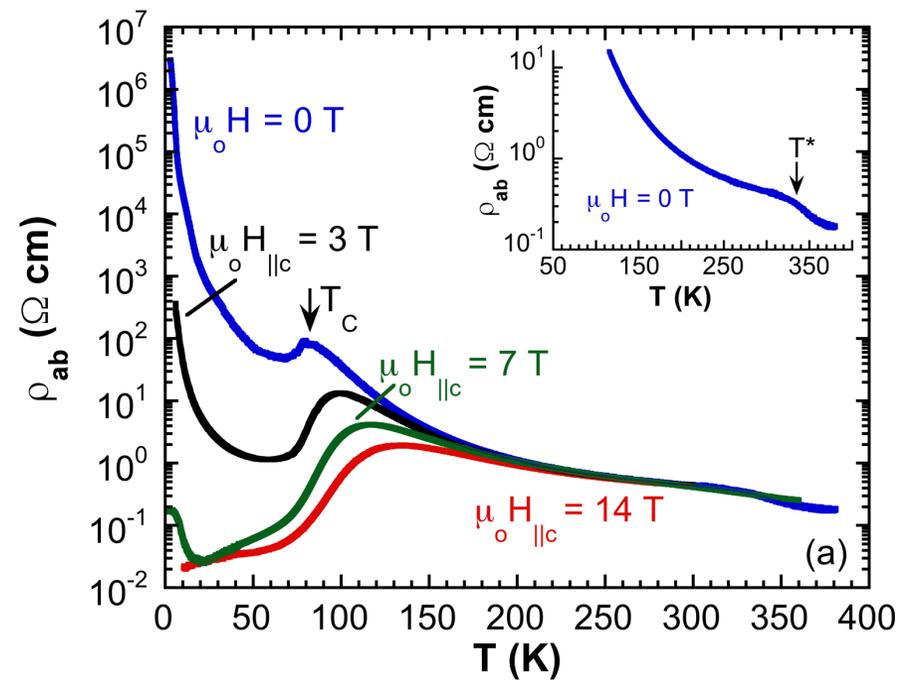

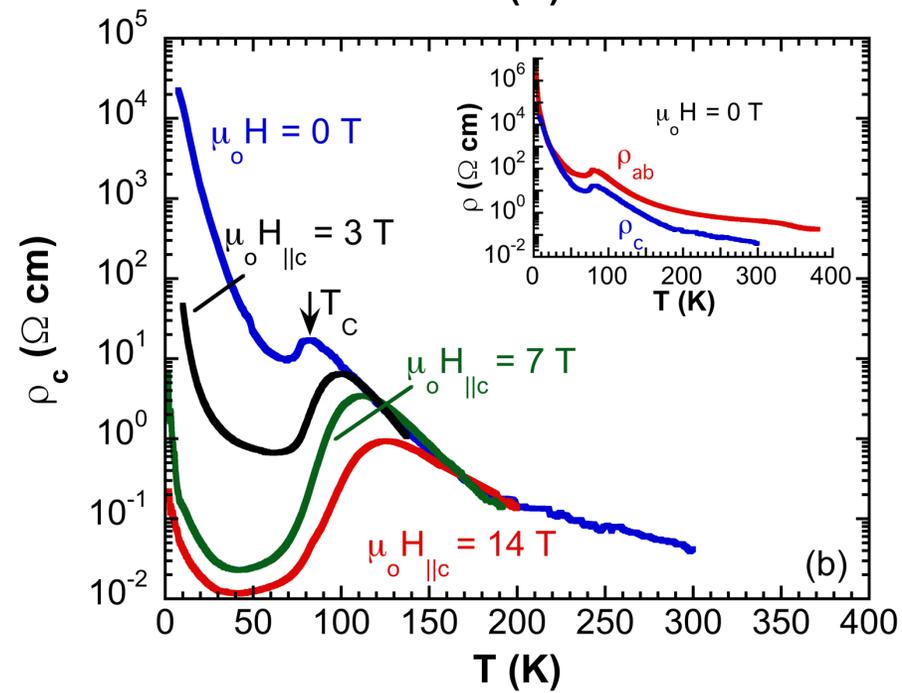

Figure 3

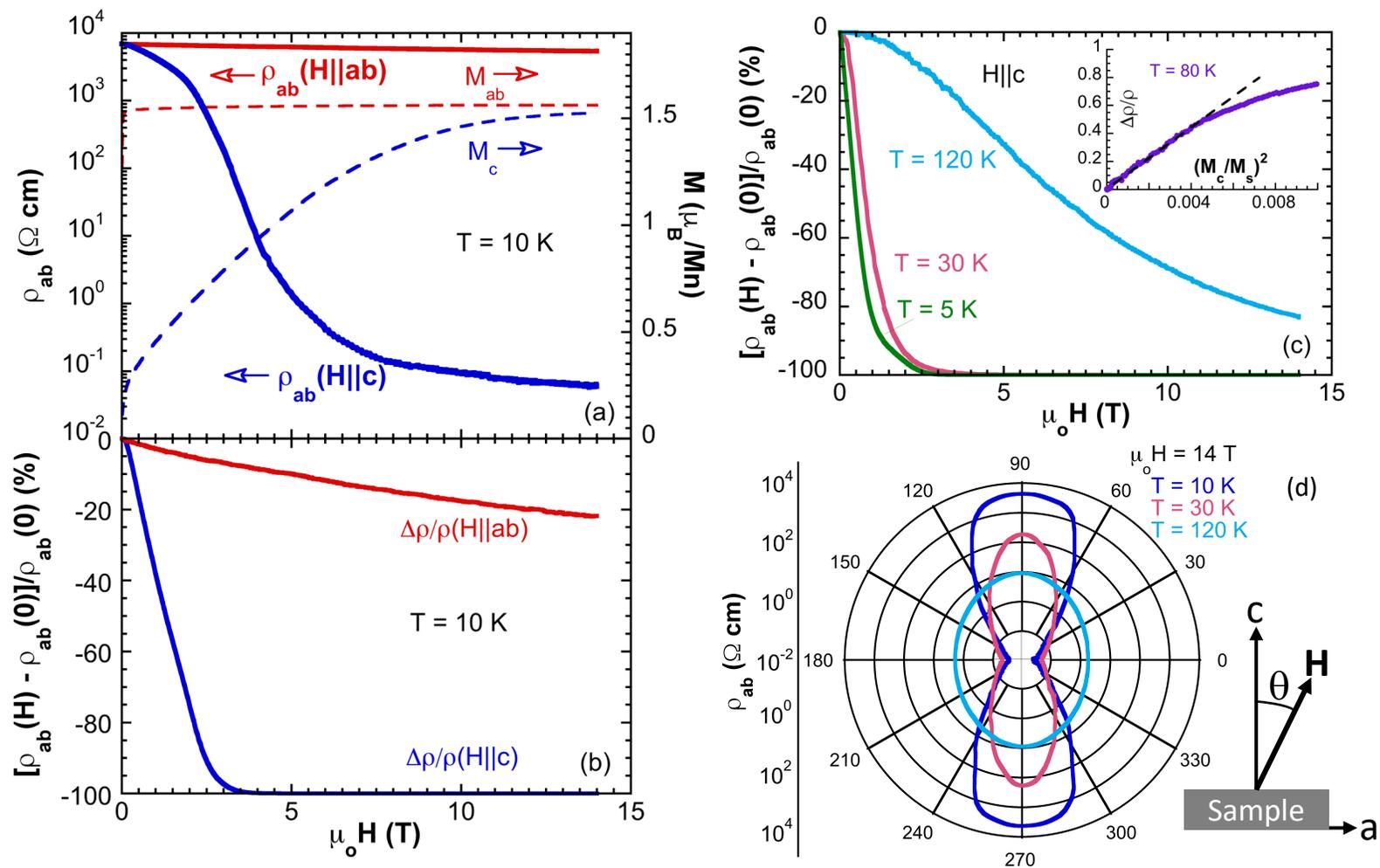

Figure 4

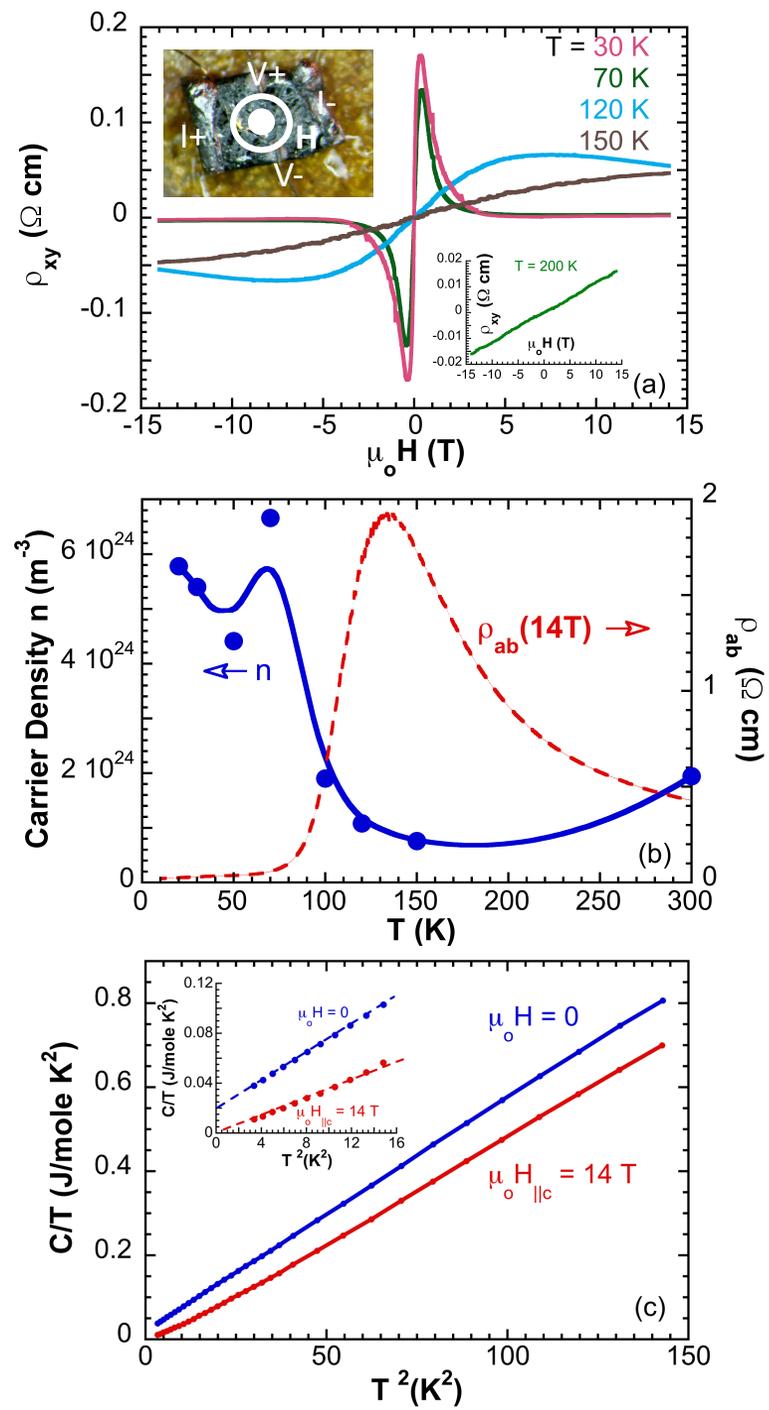

Figure 5